\documentclass[journal]{IEEEtran}

\usepackage{mathrsfs,stackrel}
\usepackage{amsmath}
\usepackage{amssymb}
\usepackage{amsfonts}
\usepackage{graphicx}

\newcommand{\Der}{\mathrm{d}}
\newcommand{\Img}{\mathrm{i}}
\newcommand{\Eul}{\mathrm{e}}

\newcommand{\ns}{n_{\text{s}}}
\newcommand{\nn}{n_{\text{n}}}

\newcommand{\NoisePSD}{\mathscr{N}}
\newcommand{\SigPower}{P}
\newcommand{\PTr}{\tau}
\newcommand{\EField}{\mathscr{E}}
\newcommand{\DCurr}{\mathscr{I}}

\newcommand{\PPMtime}{{\cal T}}

\ifCLASSINFOpdf
\else
\fi

\usepackage{amsmath}
\usepackage{amssymb}
\usepackage{amsfonts}
\usepackage{graphicx}

\usepackage{color}

\begin{document}

\title{Quantum Limits in Optical Communications}

\author{Konrad~Banaszek,~\IEEEmembership{Senior Member,~IEEE,}
        Ludwig Kunz, Micha{\l} Jachura, and Marcin Jarzyna
\thanks{The authors are with the Centre for Quantum Optical Technologies, Centre of New Technologies,
University of Warsaw, Banacha 2c, 02-097 Warszawa, Poland. K.B. and L.K. are also with the Faculty of Physics,
University of Warsaw, Pasteura 5, 02-093 Warszawa, Poland.}%
\thanks{Preparation of this paper was supported by the project ``Quantum Optical Technologies'' carried
out under the International Research Agendas Programme of the Foundation for
Polish Science co-financed by the European Union through the European Regional
Development Fund.}%
\thanks{Manuscript received xxx.}}

\markboth{}%
{Banaszek \MakeLowercase{\textit{et al.}}: Quantum Limits of Optical Communications}

\maketitle

\begin{abstract}
This tutorial reviews the Holevo capacity limit as a universal tool to analyze the ultimate transmission rates in a variety of optical communication scenarios, ranging from conventional optically amplified fiber links to free-space communication with power-limited optical signals. The canonical additive white Gaussian noise model is used to describe the propagation of the optical signal. The Holevo limit exceeds substantially the standard Shannon limit when the power spectral density of noise acquired in the course of propagation is small compared to the energy of a single photon at the carrier frequency per unit time-bandwidth area.
General results are illustrated with a discussion of efficient communication strategies in the photon-starved regime.
\end{abstract}


\IEEEpeerreviewmaketitle

\section{Introduction}

It has been recognized for a long time that quantum effects set limits on the information capacity of optical communication links \cite{GordonProcIRE1962}. The simplest argument is that detection of light based on the photoelectric effect is inherently noisy. The lowest attainable noise level---usually referred to as the shot noise level---can be determined from the quantum mechanical description of the photodetection process \cite{MandelWolfSemiclPhot}. The resulting Poisson channel model is directly applicable to intensity modulation-direct detection communication systems \cite{ShamaiIEEProc1990}. Shot noise of the photodetection process determines also the best attainable precision of measuring quadratures of the electromagnetic field by means of homodyning or heterodyning \cite{ShapiroIEEEJQE1985} which are used as detection techniques in coherent communications \cite{SalzIEEECOMM1986,KikuchiJLT2016}.

The above argument assumes that information is encoded in a well-defined classical property of the electromagnetic field such as the intensity or the phase. However, one can adopt a more fundamental quantum mechanical perspective on optical  communication \cite{Shapiro2009}. In general, the information to be transmitted is carried by certain quantum states of the electromagnetic field. These states should be discriminated by the receiver in a way that maximizes the information rate. The receivers can implement unconventional detection strategies that exhibit sensitivity beyond shot-noise-level direct detection or coherent detection \cite{MuellerUsugaNJP2012,ChenHabifNPH2012,BecerraFanNatPhot2013}.
Another possibility is to use non-classical states of light for communication, such as Fock states, that carry a well-defined number of photons \cite{GordonProcIRE1962,YuenOzawaPRL1993,CavesDrummondRMP1994}, or squeezed states, that exhibit quadrature fluctuations below the shot noise level \cite{YuenShapiro1IEEETIT1978,YamamotoHausRMP1986}. In order to identify the ultimate quantum limit of an optical communication link, one should carry out optimization over all physically permitted measurement strategies \cite{DaviesIEEETIT1978} and all ensembles of input quantum states used to carry information under relevant physical constraints, such as a restriction on the average power of the optical signal. Impressively, theoretical developments in quantum information science have provided tools to derive the ultimate quantum capacity limits in a closed analytical form for common models of optical communication links. The basic tool is Holevo's theorem \cite{Holevo1973}, which provides a tight bound on the mutual information attainable for a given ensemble of input quantum states \cite{Hausladen1996,Schumacher1997,Holevo1998}. For a scenario when a propagating optical signal experiences linear attenuation or amplification and acquires a random additive white Gaussian noise (AWGN) component, a rigorous proof of the quantum capacity limit has been presented recently \cite{GiovannettiGarciaPatronNPH2014} following earlier conjectures \cite{HolevoWernerPRA2001,GiovannettiGuhaPRA2004,GarciaPatronPRL2012}.

The purpose of this paper is to provide an introduction to the Holevo capacity limit and to relate it to the standard Shannon capacity limit for linear AWGN channels used as a benchmark when evaluating the performance of optical communication systems \cite{KahnHoJLT2004,EssiambreKramerJLT2010,WinzerJLT2012,BayvelMaherPTRSA2016}. When discussing quantum capacity limits it is essential to distinguish between noise contributed by the propagation of the optical signal and that introduced by the detection process. As this tutorial will emphasize, there is no single universal figure for the detection noise, which needs to be characterized specifically for a given detection scheme.
For clarity, the contribution from the noisy propagation of an optical signal will be referred to as the excess noise. In contrast to the Shannon capacity limit, which is customarily expressed in terms of the signal-to-noise ratio, the Holevo capacity limit uses an absolute scale for the signal and the excess noise strengths defined by the energy of a single photon at the signal carrier frequency. Only when the power spectral density of the excess noise exceeds this energy per unit time-bandwidth area, the Holevo capacity limit effectively coincides with its Shannon counterpart. This tutorial will illustrate the gap between the Holevo and the Shannon capacity limits using the example of photon-starved communication, which provides an interesting use case to develop unconventional detection strategies.

The paper starts with a mathematical description of the optical signal and its propagation in Sec.~\ref{Sec:OpticalSignal}. Shot noise level in conventional detection techniques is discussed in Sec.~\ref{Sec:Detection}. Sec.~\ref{Sec:Shannon} reviews the standard Shannon capacity limit paying attention to distinction between the excess noise and the detection noise. Sec.~\ref{Sec:Holevo} introduces the Holevo capacity limit and identifies the regime where it can be related directly to the Shannon formula.
Efficiency limits of photon-starved communication are discussed in Sec.~\ref{Sec:PhotonStarved} with examples of unconventional detection strategies given in Sec.~\ref{Sec:Multisymbol}. Finally, Sec.~\ref{Sec:Conclusions} concludes the paper.

\section{Optical signal}
\label{Sec:OpticalSignal}

\begin{figure}[t]
	\begin{center}
		\includegraphics[scale=1.05]{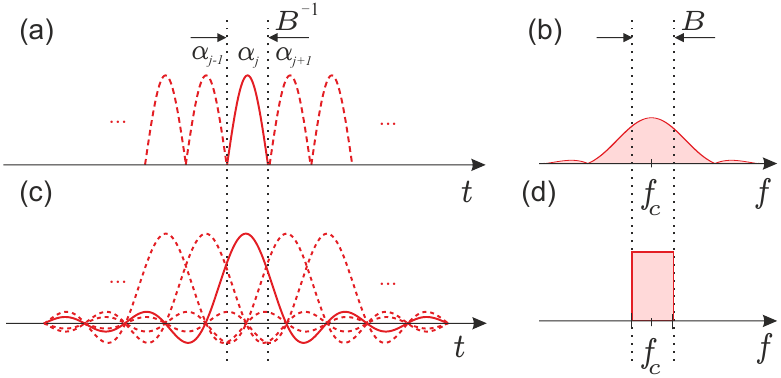}
		\caption{(a) A schematic representation of an optical signal composed of pulses with complex amplitudes $\ldots, \alpha_{j-1}, \alpha_{j}, \alpha_{j+1}, \ldots$ occupying slots of duration $B^{-1}$. (b) The slot rate $B$ characterizes the extent of the spectrum. Generally, the signal spectral support is larger than $B$
\cite{EssiambreKramerJLT2010}. (c) Pulses described by the sinc profile specified in Eq.~(\ref{Eq:sinc}) overlap in time, but satisfy the orthogonality condition (\ref{Eq:ModeNormalization}). (d) The bandwidth occupied by a sinc pulse train is equal to the slot rate $B$.} \label{Fig:Signal}
	\end{center}
\end{figure}

We will consider a narrowband, linearly polarized optical signal in the form of uniformly spaced pulses (wavepackets) located in temporal slots of duration $B^{-1}$, depicted schematically in Fig.~\ref{Fig:Signal}(a). The parameter $B$ will be referred to as the slot rate. A single pulse is described by a normalized complex profile $u(s)$ parameterized with a dimensionless time $s$ that satisfies the orthogonality condition
\begin{equation}
\label{Eq:ModeNormalization}
\int_{-\infty}^{\infty} \Der s \, u^\ast(s-j) u(s) = \delta_{j0}
\end{equation}
with its replica displaced by any integer number $j$ of slots.
The electric field $E(t)$ of the optical signal can be written as
\begin{equation}
E(t) = \Eul^{ - 2 \pi \Img f_{\text{c}} t} \EField(t)  + \Eul^{ 2 \pi \Img f_{\text{c}} t}  \EField^\ast (t),
\end{equation}
where $f_{\text{c}}$ is the carrier frequency and $\EField(t)$ is the complex analytic signal envelope given  by
\begin{equation}
\label{Eq:Efield}
\EField(t) = \sqrt{\frac{h f_{\text{c}}}{2\epsilon A_{\text{eff}}}} \sum_{j=-\infty}^{\infty} \alpha_{j} u_j(t), \qquad u_j(t) = \sqrt{B} u(Bt-j).
\end{equation}
Here $h = 6.626 \times 10^{-34}\;\text{J}\cdot \text{s}$ is Planck's constant, $\epsilon$ is the permittivity of the propagation medium, $A _{\text{eff}}$ is the effective area of the transverse spatial mode in which the signal propagates, and $\alpha_j$ are the complex amplitudes of individual wavepackets.
The normalization factor in Eq.~(\ref{Eq:Efield}) is chosen such that the average optical power carried by the signal can be expressed with the help of Eq.~(\ref{Eq:ModeNormalization}) as:
\begin{equation}
\SigPower =  \lim_{T\rightarrow \infty}\frac{1}{T} \int_{-T/2}^{T/2} \Der t \int_{A_{\text{eff}}} \Der^2 {\bf r}  \, 2\epsilon |{\EField}(t)|^2 = B h f_{\text{c}} {\mathbb E}[|\alpha_j|^2].
\end{equation}
The squared absolute value $|\alpha_j|^2$ has the interpretation of the mean photon number carried by the $j$th pulse and the expectation value
\begin{equation}
\label{Eq:nbar}
\bar{n} = {\mathbb E}[|\alpha_j|^2] = \frac{\SigPower}{B h f_{\text{c}}}
\end{equation}
is the average signal photon number per temporal slot. The amplitudes $\alpha_j$ are usually drawn from a discrete set that can be visualized as a constellation in the complex parameter plane. Individual points in the constellation are referred to as symbols. While practical communication is predominantly based on discrete constellations, analysis of capacity limits should include general, possibly continuous probability distributions for the complex amplitudes $\alpha_j$.

A narrowband scenario with $B \ll f_{\text{c}}$ will be considered here.
The normalized spectrum of the signal is given by $\bigl|\tilde{u}\bigl( (f-f_{\text{c}})/B\bigr)\bigr|^2/B$, where $\tilde{u}(\nu)=\int_{-\infty}^{\infty} \Der s\, \Eul^{2\pi \Img \nu s}u(s)$ is the Fourier transform of the pulse profile $u(s)$. As shown in Fig.~\ref{Fig:Signal}(b), the slot rate $B$ characterizes the extent of the signal spectrum in the frequency domain \cite{EssiambreKramerJLT2010}.
The formalism used here includes also the case of wavepackets overlapping in the temporal domain, such as the commonly used sinc profile illustrated with Fig.~\ref{Fig:Signal}(c)
\begin{equation}
u(s) = \frac{\sin(\pi s)}{\pi s}.
\label{Eq:sinc}
\end{equation}
In this particular case the signal spectrum has a rectangular form depicted in Fig.~\ref{Fig:Signal}(d) extending from $f_{\text{c}} - B/2$ to $f_{\text{c}} + B/2$ and the slot rate $B$ has direct interpretation of the signal bandwidth.

\begin{figure}
	\begin{center}
		\includegraphics[scale=1]{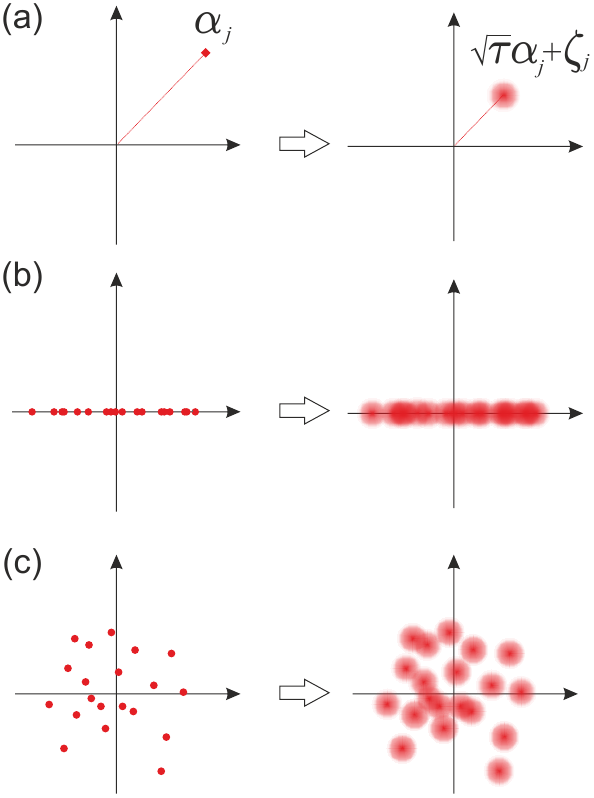}
		\caption{(a) Transformation of complex amplitudes $\alpha_{j}$ in the linear additive white Gaussian noise model.  (b) One-dimensional Gaussian ensemble. (c) Two-dimensional Gaussian ensemble.} \label{Fig:ChannelAction}
	\end{center}
\end{figure}

The propagation of the optical signal through the physical medium will be described using the standard AWGN model, in which complex amplitudes $\alpha_j$ of individual pulses undergo a transformation
\begin{equation}
\label{Eq:alphajchannel}
\alpha_j \rightarrow \alpha_j' = \sqrt{\PTr} \alpha_j + \zeta_j.
\end{equation}
Here the transmission coefficient $\PTr \ge 0$ specifies the change in the optical signal power in the course of propagation and $\zeta_j$ are  random variables that characterize noise added in individual slots. It is important to stress that these variables do not include the noise contributed by the detection process, which will be treated separately. 
For clarity, the field component contributed by the variables $\zeta_j$ will be referred to as the excess noise. In the AWGN model for the excess noise, $\zeta_j$ are mutually independent complex-valued Gaussian random variables $\zeta_j \sim {\cal CN}(0,\nn)$ with zero mean and the variances of their real and imaginary parts equal to
\begin{equation}
\label{Eq:RexiImxi}
\mathbb{E}[(\text{Re} \zeta_j)^2] = \mathbb{E}[(\text{Im} \zeta_j)^2] = \nn/2.
\end{equation}
The total variance $\nn = \text{Var}[\zeta_j]$ can be interpreted as the mean number of excess noise photons added per one temporal slot. In the white noise scenario, $\nn$ is independent of the slot rate $B$ and can be expressed as
\begin{equation}
\nn = \frac{\NoisePSD}{hf_\text{c}},
\end{equation}
where $\NoisePSD$ is the excess noise power spectral density. In order to keep the notation concise, the average received signal photon number per slot will be denoted as
\begin{equation}
\label{Eq:ns}
\ns = \PTr \bar{n} = \frac{\PTr \SigPower}{hf_\text{c}}.
\end{equation}
When the signal power is attenuated, i.e.\ $\PTr < 1$, the parameter $\nn$ can assume any nonnegative value. In this regime, loss-only propagation is defined by $\nn = 0$. However, when the output signal emerges amplified, i.e.\ $\PTr > 1$, the excess noise must be added in the amount of at least $\nn \ge \PTr -1$. This requirement can be interpreted within the quantum theory of optical amplification as a consequence of the Heisenberg uncertainty principle \cite{HausMullenPR1962,Caves1982}.

\section{Conventional detection}
\label{Sec:Detection}

\begin{figure}
	\begin{center}
		\includegraphics[scale=1]{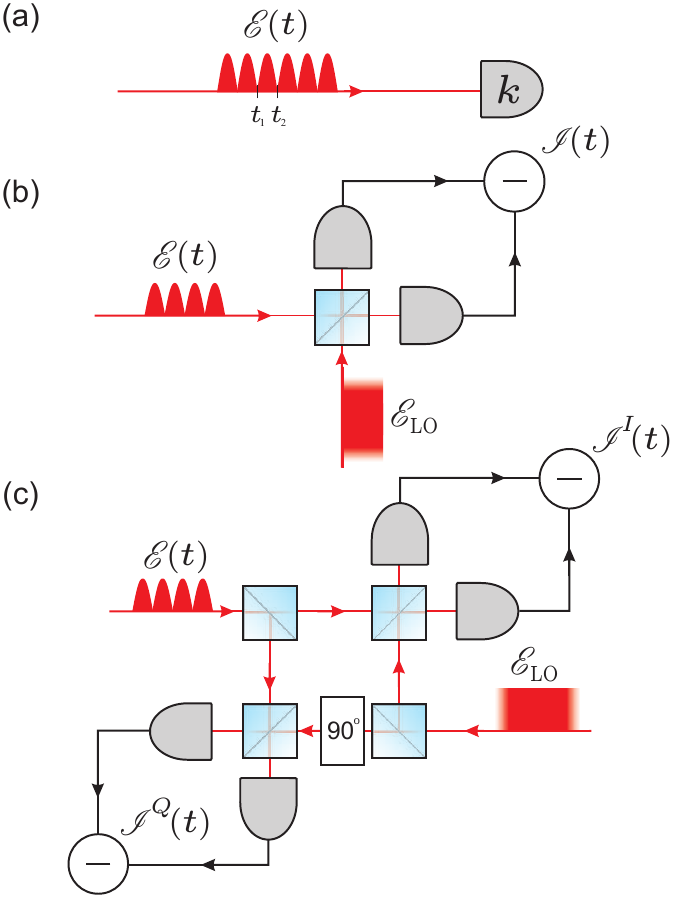}
		\caption{(a) Idealized direct detection of an optical field $\EField(t)$. The measurement outcome over an interval lasting from $t_1$ to $t_2$ is a discrete number of $k$ photocounts. (b) Balanced homodyne detection of one field quadrature using a continuous wave local oscillator with a complex amplitude $\EField_{\text{LO}}$. (c) Measurement of both $I$ and $Q$ quadratures using two balanced homodyne setups with the local oscillator phases set respectively to $0^\circ$ and $90^\circ$.}
\label{Fig:Detection}
	\end{center}
\end{figure}

Standard methods to measure the received optical signal are direct detection and homodyne detection of one or both quadratures of the electromagnetic field, shown in Fig.~\ref{Fig:Detection}.
This section will briefly review statistical properties of these measurements assuming that the photodetection process is free from technical imperfections and operates at the shot noise level. The objective is to identify the minimum amount of noise that has to occur in the readout of an optical signal by conventional detection methods.

The standard techniques to measure an optical field are based on the photoelectric effect, when incident light ejects electrons from a photocathode or generates electron-hole pairs in a semiconductor. At the fundamental level the number of produced elementary photocarriers is integer, and with a sufficiently low-noise gain mechanism it can be read out from the photodetection device in such discrete form as the photocount number \cite{MandelProcPhysSoc1959}. Within the framework of the quantum theory of electromagnetic radiation, generation of each photocarrier is associated with an absorption of a single photon from the field illuminating the photodetector  \cite{Kelley1964}. However, equivalent statistical predictions regarding the photodetection process can be obtained by treating the electromagnetic field as a classical entity and using a quantum mechanical model only to describe the charge carriers in the photodetector \cite{MandelJOSA1977}. Such a semiclassical description of the photodetection process is valid as long as one does not deal with non-classical states of light that cannot be legitimately described within classical electrodynamics.

In an idealized scenario when the photodetector has unit detection efficiency and produces no dark counts, the probability of generating $k$ photocarriers over a time interval lasting from $t_1$ to $t_2$ by an incident electromagnetic field with a complex envelope $\EField(t)$ is given by the Poissonian statistics
\begin{equation}
\label{Eq:pkDD}
p_k = \exp(-\bar{k}) \frac{{\bar{k}}^{k}}{k!},
\end{equation}
where the expectation value of the photocount number
\begin{equation}
\bar{k} = \mathbb{E}[k] = \frac{2\epsilon }{h f_{\text{c}}} \int_{A_{\text{eff}}} \Der^2 {\bf r} \int_{t_1}^{t_2} \Der t
 \, |\EField(t)|^2.
\end{equation}
can be interpreted as the mean number of photons carried by the field within the measurement interval. The variance of the photocount number $\text{Var}[k]=\bar{k}$ is referred to as the shot noise level.
Taking $\EField(t)$ in the form given by Eq.~(\ref{Eq:Efield}), when the integration interval contains only the $j$th signal pulse $u_j(t)$ and $\int_{t_1}^{t_2} \Der t
\, |u_j(t)|^2 =1$, the mean photocount number $\bar{k}$ can be identified with the mean number of photons in that
pulse, $\bar{k} = |\alpha_j|^2$.

Direct detection reveals information only about the intensity of the incident electromagnetic field. A standard phase-sensitive measurement technique is homodyning, where the incoming signal is superposed on a beam splitter with an auxiliary local oscillator (LO) beam that has the same frequency as the signal carrier. We will consider a model of a homodyne setup shown in Fig.~\ref{Fig:Detection}(b), where LO is prepared as a continuous wave field with a complex amplitude $\EField_{\text{LO}}= |\EField_{\text{LO}}|\Eul^{\Img \phi}$. Both the signal and LO fields are superposed on a balanced 50:50 beam splitter \cite{YuenChanOL1983} whose output ports are monitored by idealized photodetectors producing photocount statistics described by Eq.~(\ref{Eq:pkDD}). The two fields leaving the beam splitter are described by $\bigl(\EField(t)  \pm \EField_{\text{LO}}\bigr)/\sqrt{2}$. Let us divide the time axis into discrete intervals of duration $\Delta t$ indexed using an integer $i=\ldots, -1,0,1,\ldots$, with the $i$th interval centered at $t_i = i\Delta t$. When the signal and the LO fields do not fluctuate, the joint probability distribution of registering $k_{i+}$ and $k_{i-}$ photocounts on the two photodetectors over the $i$th interval is a product of two Poissonian distributions with respective expectation values
\begin{align}
\bar{k}_{i\pm} =
\mathbb{E}[k_{i\pm}]
& =  \frac{\epsilon }{h f_{\text{c}}} \int_{A_{\text{eff}}} \Der^2 {\bf r}
\int_{t_i - \Delta t/2}^{t_i + \Delta t/2} \Der t \,
| \EField(t)  \pm \EField_{\text{LO}}|^2  \nonumber \\
& \approx  \frac{\epsilon A_{\text{eff}}}{h f_{\text{c}}} | \EField(t_i)  \pm \EField_{\text{LO}}|^2 \Delta t,
\end{align}
where the second approximate expression holds if $\Delta t$ is shorter than the temporal variation of the signal field envelope $\EField(t)$. If the LO field carries a macroscopic number of photons over the integration time $\Delta t$, i.e.\ $\epsilon |\EField_{\text{LO}}|^2 \Delta t \gg h f_{\text{c}}$, the photocount numbers ${k}_{i\pm}$ can be treated  as continuous variables \cite{YuenShapiro3IEEEIT1980} characterized by normal distributions ${k}_{i\pm} \sim {\cal N}(\bar{k}_{i\pm}, \bar{k}_{i\pm})$ that approximate Poisson distributions for ${k}_{i\pm} \gg 1$. Consider now the rescaled differential photocurrent
\begin{equation}
\label{Eq:yidef}
\DCurr_i = \frac{1}{2|\EField_{\text{LO}}|\Delta t} \sqrt{\frac{h f_{\text{c}}}{\epsilon A_{\text{eff}}}} ( {k}_{i+}-{k}_{i-} ).
\end{equation}
Under present assumptions the differential photocurrent is a Gaussian random variable with the
 expectation value
\begin{equation}
\label{Eq:Expyi}
\mathbb{E}[\DCurr_i] = 2 \sqrt{\frac{\epsilon A_{\text{eff}}}{h f_{\text{c}}}} \text{Re} [\Eul^{-\Img \phi}
\EField (t_i)],
\end{equation}
and the variance given by
\begin{equation}
\text{Var} [\DCurr_i] = \frac{h f_{\text{c}}}{4 \epsilon A_{\text{eff}} |\EField_{\text{LO}} |^2  (\Delta t)^2 }
\{ \text{Var}[{k}_{i+}] + \text{Var}[{k}_{i-}] \} \approx \frac{1}{2 \Delta t},
\end{equation}
where  only the leading-order terms in $|\EField_{\text{LO}}|$ have been retained when evaluating $\text{Var}[{k}_{i\pm}]$. Provided that the signal and the LO fields do not exhibit any fluctuations, the differential photocurrent noise is uncorrelated between different time intervals, $\text{Cov}[\DCurr_i, \DCurr_{i'}]=0$ for $i\neq {i'}$. In the remainder, it will be convenient to apply the limiting transition $\Delta t \rightarrow 0$ and treat $t$ as a coarse-grained time variable. In this limit
\begin{equation}
\label{Eq:ExpDcurrt}
\mathbb{E}[\DCurr(t)] = 2 \sqrt{\frac{\epsilon A_{\text{eff}}}{h f_{\text{c}}}} \text{Re} [\Eul^{-\Img \phi}
\EField (t)],
\end{equation}
and
\begin{equation}
\label{Eq:CovDCurr}
\text{Cov}[\DCurr(t), \DCurr(t')] = \frac{1}{2} \delta(t-t').
\end{equation}
Applying to $\DCurr(t)$ a filter function \cite{MecozziEssiambreJLT2012} described in the temporal domain by a normalized real profile $v(t)$ yields a quadrature variable
\begin{equation}
y = \int_{-\infty}^{\infty} \Der t \,  v(t) \DCurr(t).
\end{equation}
The expectation value of $y$ can be directly calculated using Eq.~(\ref{Eq:ExpDcurrt})
for $\EField(t)$ given by Eq.~(\ref{Eq:Efield}) to be equal to
\begin{equation}
\label{Eq:Expy}
\mathbb{E}[y] = \sqrt{2} \sum_{j=-\infty}^{\infty} \text{Re} \left(\Eul^{-\Img \phi} \alpha_j
\int_{-\infty}^{\infty} \Der t \, v(t)  u_j(t)
\right),
\end{equation}
while Eq.~(\ref{Eq:CovDCurr}) gives variance
\begin{equation}
\label{Eq:Vary}
\text{Var}[y]=\int_{-\infty}^{\infty} \Der t \int_{-\infty}^{\infty} \Der {t'} \,
\text{Cov}[\DCurr(t), \DCurr(t')] = \frac{1}{2}.
\end{equation}
When the filter  matches the profile of the $j$th signal wavepacket, $v(t) = u_j(t)$, Eq.~(\ref{Eq:Expy}) reduces to $\mathbb{E}[y] = \sqrt{2}\text{Re} (\Eul^{-\Img \phi} \alpha_j)$, which follows from the orthogonality condition (\ref{Eq:ModeNormalization}). This requires that the wavepacket profile is real. By selecting the LO phase $\phi^I=0^\circ$ or $\phi^Q=90^\circ$ one can detect respectively either the $I$ or the $Q$ field quadrature. Eqs.~(\ref{Eq:Expy}) and (\ref{Eq:Vary}) imply that the measurement outcome $y^{I,Q}$ is characterized by a Gaussian probability distribution
\begin{equation}
\label{Eq:onequadrature}
p(y^{I,Q}) = \frac{1}{\sqrt{\pi}} \exp \{- [y^{I,Q} - \sqrt{2} \text{Re} (\Eul^{-\Img \phi^{I,Q}}\alpha_j)]^2\}.
\end{equation}
Note that the variance of this distribution stems from the shot noise in the photodetection process and its numerical value $\text{Var}[y^{I,Q}]= 1/2$ is determined by the rescaling of the differential photocurrent used in Eq.~(\ref{Eq:yidef}).
Remarkably, quadrature distributions have been recently measured  at the shot-noise-level for binary phase shift keyed (BPSK) signals sent from a geostationary satellite to an optical ground station equipped with a homodyne receiver \cite{GuntKhanOPT2017}.

A way to measure both $I$ and $Q$ quadratures for a single pulse is to split the input signal equally between two homodyne setups and to use LO with phases $\phi^I = 0^\circ$ and $\phi^Q = 90^\circ$ \cite{WalkerCarrollOQE1986}. This arrangement, known in the context of optical communication as phase diversity homodyne detection \cite{NoeIEEEPTL2005,Ly-GagnonJLT2006,KikuchiIEEEJSTQE2006}, is shown in Fig.~\ref{Fig:Detection}(c). The rescaled differential photocurrents $\DCurr^I(t)$ and $\DCurr^Q(t)$  are defined analogously to Eq.~(\ref{Eq:yidef}) using the LO amplitude fed into an individual homodyne setup
and taken in the limit  $\Delta t \rightarrow 0$. Their expectation values read:
\begin{align}
\mathbb{E}[\DCurr^I(t)] & = \sqrt{\frac{2\epsilon A_{\text{eff}}}{h f_{\text{c}}}}
\text{Re} [\EField (t)], \nonumber \\
\mathbb{E}[\DCurr^Q(t)] & = \sqrt{\frac{2\epsilon A_{\text{eff}}}{h f_{\text{c}}}}
\text{Im} [\EField (t)].
\end{align}
Note the reduction by a factor $\sqrt{2}$ compared to Eq.~(\ref{Eq:ExpDcurrt}), as each homodyne setup receives only half of the input signal power. Differential photocurrent noise is characterized by
$\text{Cov}[\DCurr^I(t), \DCurr^I(t')]  = \text{Cov}[\DCurr^Q(t), \DCurr^Q(t')] = \frac{1}{2} \delta(t-t')$
and $\text{Cov}[\DCurr^I(t), \DCurr^Q(t')] = 0$. In the case of a two-quadrature measurement one can take a complex normalized filter function $v(t)$ and define
\begin{equation}
y^I + \Img y^Q = \int_{-\infty}^{\infty} \Der t \,  v^\ast(t) [\DCurr^I(t)+ \Img \DCurr^Q(t)].
\end{equation}
When $\EField(t)$ has the form given in Eq.~(\ref{Eq:Efield}) one obtains
\begin{equation}
\mathbb{E}[y^I + \Img y^Q] =  \sum_{j=-\infty}^{\infty}  \alpha_j
\int_{-\infty}^{\infty} \Der t \, v^\ast(t) u_j(t) .
\end{equation}
and $\text{Var}[y^I]= \text{Var}[y^Q]=1/2$. If $v(t)$ matches the profile $u_j(t)$ of the $j$th wavepacket,
the joint probability distribution for $y^I$ and $y^Q$ can be compactly written as:
\begin{equation}
\label{Eq:pyIyQ}
p(y^I, y^Q) = \frac{1}{\pi}
\exp[ - (y^I  - \text{Re} \alpha_j)^2 - ( y^Q -  \text{Im}\alpha_j)^2].
\end{equation}
Note that in the present case the profile $u_j(t)$ can be complex.

Compared to the one-quadrature measurement described by Eq.~(\ref{Eq:onequadrature}), the complex  amplitude $\alpha_j$ in Eq.~(\ref{Eq:pyIyQ}) is reduced by a factor $\sqrt{2}$ that stems from dividing the signal power between two homodyne setups, while the variances of individual outcomes $y^I$ and $y^Q$ remains at the same level, $\text{Var}[y^I]=\text{Var}[y^Q]=1/2$.
In the quantum theory of electromagnetic radiation $I$ and $Q$ quadratures are described by non-commuting observables. 
Simultaneous measurement of such observables on a single quantum system has to be accompanied by additional uncertainty \cite{ArthursKellyBSTJ1965,WodkiewiczPRL1984}.
This can be viewed as the fundamental reason for the reduced signal-to-noise ratio when both quadratures are detected for one optical pulse.

Importantly, matched filtering allows in principle for shot-noise-level determination of quadratures for individual symbols even in the case of temporally overlapping pulses, provided that the orthogonality condition (\ref{Eq:ModeNormalization}) is satisfied.
In contrast, standard direct detection requires that the individual pulses are confined to separate slots in order to discriminate between their contributions to the photocount statistics.
This restriction can be in principle lifted using the recently developed technique of quantum pulse gating, which
allows one to demultiplex individual temporal wavepackets from an orthogonal set by carefully engineered up-conversion in a $\chi^{(2)}$ nonlinear medium \cite{BrechtReddyPRX2015,AllgaierAnsariNCOMM2017,ReddyRaymerOPTICA2018}.

\section{Communication channel}
\label{Sec:Shannon}

\begin{figure}
	\begin{center}
		\includegraphics[scale=1]{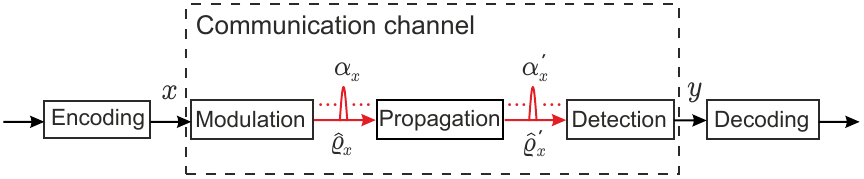}
		\caption{A generic communication scenario. The input $x$ defines modulation of the optical signal $\alpha_x$. After propagation, detection of the output signal $\alpha'_x$ produces outcome $y$. The complete quantum mechanical scenario allows for preparation of general quantum states described by density operators $\hat{\varrho}_x$ that are mapped onto output states $\hat{\varrho}_x'$.} \label{Fig:CommDiagram}
	\end{center}
\end{figure}

In a generic communication scenario shown in Fig.~\ref{Fig:CommDiagram}, the complex amplitude $\alpha$ for a pulse in a given slot is selected according to the value $x$ of an input random variable $X$ characterized by a probability distribution $p_x$.  The outcome $y$ of the measurement performed at the detection stage is a realization of a certain random variable $Y$. In the absence of memory effects the communication channel is characterized by a set of conditional probability distributions $p_{y|x}$. In the optical implementation considered here, these distributions are determined jointly by the map $x \rightarrow \alpha_x$, the transformation of the optical signal in the course of propagation, and the employed detection scheme. The amount of information about $X$ that can be recovered from $Y$ is quantified by the mutual information \cite{CoverThomas}
\begin{equation}
\label{Eq:IXY}
{\mathsf I}(X;Y) = H(Y) - H(Y|X),
\end{equation}
where $H(Y) = -\sum_y p_y \log_2 p_y$ and $H(Y|X) = -\sum_{x} p_x \sum_{y} p_{y|x} \log_2 p_{y|x}$ are respectively the marginal and the conditional entropy of the measurement results. According to Shannon's noisy-channel coding theorem \cite{ShannonBSTJ1948}, the maximum amount of information per channel use that can be communicated reliably at an arbitrarily low error rate is obtained by optimizing the mutual information ${\mathsf I}(X;Y)$ with respect to the input probability distribution $p_x$. This defines the capacity of a memoryless channel as
\begin{equation}
\mathsf{C} = \sup_{\{p_x\}} \mathsf{I} (X;Y).
\end{equation}

A standard illustration of the above concept is the derivation of the Shannon capacity limit. The basic theoretical tool is the Shannon-Hartley theorem \cite{Shannon1949a}, which states that the
capacity of an analog communication channel with a real input variable $x$ and a real output variable $y$ related through a Gaussian conditional probability distribution
\begin{equation}
\label{Eq:pxyGaussian}
p_{y|x} = \frac{1}{\sqrt{2\pi N}} \exp \left(- \frac{(y-\sqrt{\eta} x)^2}{2N} \right)
\end{equation}
under the constraint $\mathbb{E}[x^2] \le S$ is equal to
$ \mathsf{C} =  \frac{1}{2} \log_2 (1+\eta S/{N} )$
and is attained by the Gaussian input distribution $x \sim {\cal N}(0,S)$.

Consider first the case when only
one field quadrature, taken for concreteness to be the $I$ component, is used for communication.
Let the input variable $x$ define the complex field amplitude by $\text{Re}\alpha_x = x/\sqrt{2}$ and  $\text{Im}\alpha_x = 0$. The average power constraint can be expressed as
\begin{equation}
S = \mathbb{E}[x^2]= 2\mathbb{E}[(\text{Re}\alpha_x)^2] = 2 \bar{n}.
\end{equation}
The last equality follows from Eq.~(\ref{Eq:nbar}) taking into account the fact that in the present scenario
the imaginary part of the complex field amplitude is identically set to zero. The explicit form of the conditional distribution (\ref{Eq:pxyGaussian}) can be obtained in the current case by inserting the right hand side of  Eq.~(\ref{Eq:alphajchannel}) into Eq.~(\ref{Eq:onequadrature}) and averaging over the excess noise. The resulting scaling factor  $\eta=\PTr$ is simply the power transmission coefficient for the optical field.
The variance $N$ is a sum of two contributions. The first one, equal to $\mathbb{E} [(\sqrt{2}\text{Re} \zeta_j)^2] = \nn$ as implied by (\ref{Eq:RexiImxi}), stems from the excess noise, while the second one comes from the homodyne measurement itself. If the measurement is carried out at the shot noise level, the latter contribution is $1/2$ according to Eq.~(\ref{Eq:Vary}) and
$N  =  \nn +1/2$. Consequently, for single-quadrature communication one obtains the Shannon capacity limit in the form
\begin{equation}
\label{Eq:CS1}
{\sf C}_{\text{S1}} 
= \frac{1}{2} \log_2 \left( 1 + \frac{4\ns}{ 2 \nn +1} \right),
\end{equation}
where the enumerator has been expressed in terms of the average received signal photon number per slot $\ns=\PTr\bar{n}$ defined in Eq.~(\ref{Eq:ns}).

In the scenario when both $I$ and $Q$ quadratures are used for information transmission, two real variables $x^I$ and $x^Q$ are used in each slot and $\alpha_x = (x^I + \Img x^Q)/\sqrt{2}$. Given that the average optical power carried by one quadrature is now $\mathbb{E}[(\text{Re}\alpha_x)^2]= \mathbb{E}[(\text{Im}\alpha_x)^2] = \bar{n}/2$ one has $S = \mathbb{E}[(x^I)^2 ]  = 2 \mathbb{E}[(\text{Re}\alpha_x)^2] = \bar{n}$ and analogously $\mathbb{E}[(x^Q)^2] = \bar{n}$.
The scaling factor between the input variables $x^I$ and $x^Q$ and the homodyne measurement outcomes $y^I$ and $y^Q$ is $\eta = \PTr/2$ as in the present case only half of the input power is directed to each of the two homodyne setups. For the same reason, only half of the excess noise power should be accounted for in the variance $N=\nn/2 +1/2$.
The Shannon capacity in bits per slot for two-quadrature communication is a sum of two equal contributions from $I$ and $Q$ components and reads \cite{HallPRA1994}
\begin{equation}
\label{Eq:CS2}
{\sf C}_{\text{S2}} = \log_2 \left( 1 + \frac{\ns}{\nn + 1} \right).
\end{equation}

\begin{figure}
	\includegraphics[width=\linewidth]{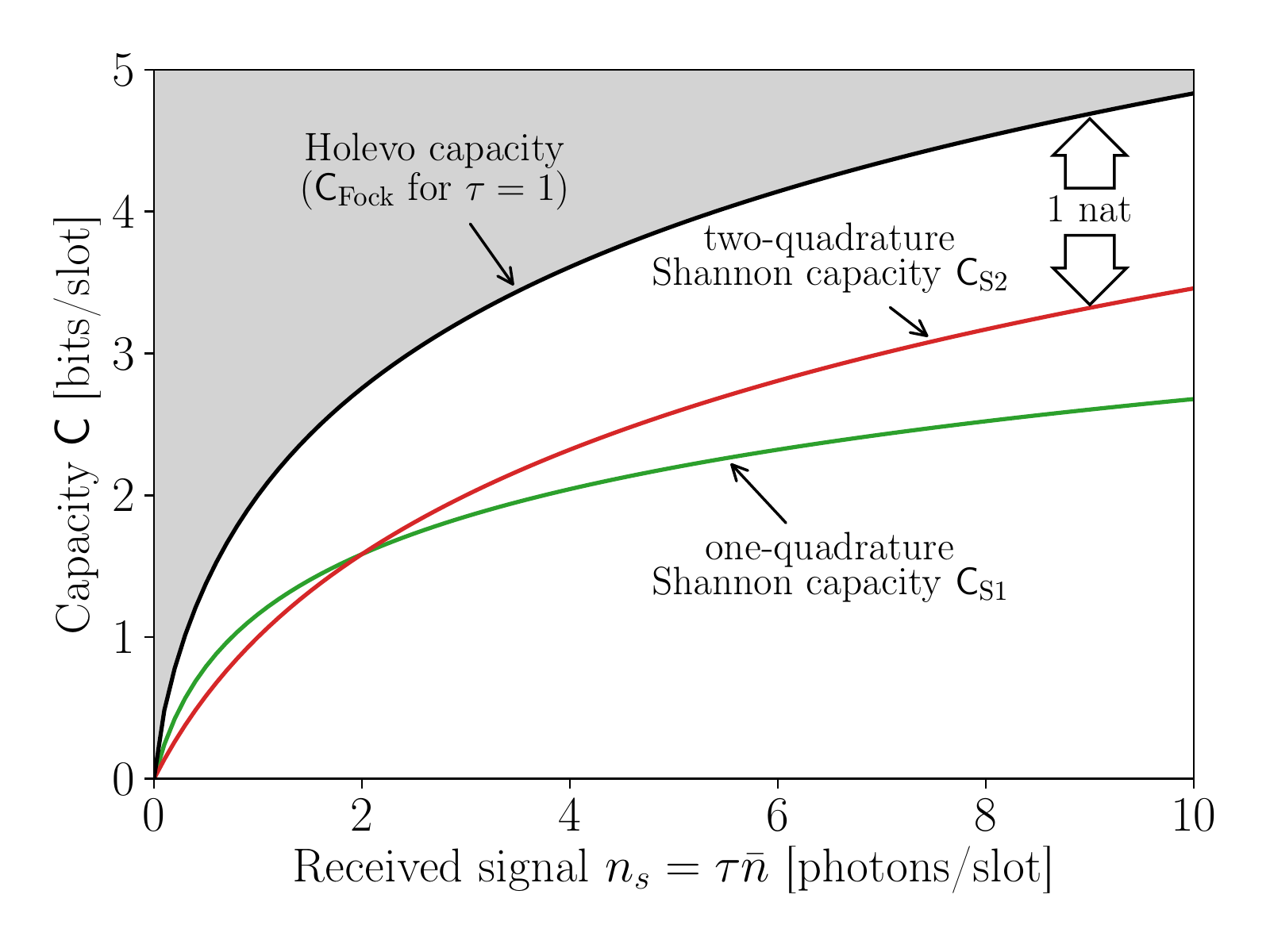}
\caption{Shannon capacities for one-quadrature $\mathsf{C}_{\text{S1}}$ and two-quadrature $\mathsf{C}_{\text{S2}}$ communication with shot noise level homodyning compared to the Holevo capacity $\mathsf{C}_{\text{H}}$ for loss-only propagation with zero excess noise. The Holevo capacity coincides with the value $\mathsf{C}_{\text{Fock}}$ attainable using non-classical Fock (photon number) states over a lossless channel.}
\label{Fig:Capacities}
\end{figure}

Fig.~\ref{Fig:Capacities} compares Shannon capacities for one- and two-quadrature encodings
as a function of the average received signal photon number $\ns$ for loss-only propagation, when the excess noise is zero $\nn=0$, and quadratures are measured at the shot noise level. It is seen that below $\ns \lesssim 2$ it is beneficial to use single-quadrature encoding, which however requires a local oscillator phase-locked to the received signal. Around $\ns \approx 2$ the capacity is in principle optimized by time-sharing between one- and two-quadrature communication \cite{Takeoka2014}, but the advantage of this strategy is minuscule.

When excess noise dominates the homodyne shot noise, $\nn \gg 1$, one can neglect the latter and
and write the Shannon capacity limit in terms of the signal-to-noise ratio (SNR) $\ns/\nn = \PTr \SigPower/(B\NoisePSD)$.
In this case, a straightforward comparison of one- and two-quadrature Shannon capacity limits yields 
\begin{equation}
{\sf C}_{\text{S1}} \approx \frac{1}{2} \log_2 \left( 1 + 2\frac{\ns}{\nn} \right) <
\log_2 \left( 1 + \frac{\ns}{\nn} \right) \approx {\sf C}_{\text{S2}},
\end{equation}
and for any SNR value it is beneficial to use both quadratures for communication. Let us note that the transmission coefficient $\tau$ and the noise density $\NoisePSD$ may incorporate respectively the non-unit detection efficiency and the excess noise contributed by the detection process.

The maximum attainable transmission rate ${\sf R}$ in bits per unit time for a given communication scenario is given by
\begin{equation}\label{Eq:InfRate}
{\mathsf R} = B \cdot {\mathsf C},
\end{equation}
where ${\sf C}$ is the corresponding capacity limit expressed in bits per slot.

\section{Holevo limit}
\label{Sec:Holevo}

The Shannon capacity limit reviewed in the preceding section relies on two 
assumptions. The first one is that the optical field carrying information can be described within the classical theory of electromagnetic radiation using the expression given in Eq.~(\ref{Eq:Efield}). The second one is that the optical signal is detected by means of homodyning, capable of measuring one or both quadratures with shot-noise-level precision. In the case of a lossless channel with unit transmission, $\PTr=1$, and no excess noise, $\nn=0$, there exists a very simple optical communication scenario suggested by Gordon \cite{GordonProcIRE1962} which beats the Shannon limit. Quantum mechanics permits preparation of a light pulse in a state which contains exactly $n$ photons, called a photon number state or a Fock state \cite{LeonhardtCUP2009}. In recent years, impressive progress in generation of such states
has been made in the context of prospective applications in quantum information processing and communication \cite{LvovskyHansenPRL2001,WaksDiamantiNJP2006,CooperWrightOpEx2013}. Suppose that the message to be transmitted is encoded in Fock states, with the $n$-photon Fock state sent with a probability $p_n$. Because a pulse prepared in an $n$-photon Fock state generates exactly $n$ counts on an ideal photodetector with unit detection efficiency and no dark counts, the Fock state transmitted over a lossless channel can be in principle identified unambiguously by direct detection. For such a communication scenario the mutual information reads
${\sf I}  = -\sum_{n=0}^{\infty} p_n \log_2 {p_n}$.
In order to identify the capacity $\mathsf{C}_{\text{Fock}}$ in this communication scenario, the above expression needs to be maximized over the probabilities $p_n \ge 0$ under the average power constraint $\sum_{n=0}^{\infty} n p_n = \bar{n}$. This task is a simple exercise in the method of Lagrange multipliers with the result $\mathsf{C}_{\text{Fock}} = g(\bar{n})$, where
\begin{equation}
\label{Eq:g(x)}
 g(\upsilon) = (\upsilon+1) \log_2 (\upsilon+1) - \upsilon \log_2 \upsilon.
\end{equation}
A graph of ${\mathsf C}_{\text{Fock}} = g(\bar{n})$ depicted in Fig.~\ref{Fig:Capacities} shows that communication with Fock states over a lossless channel exceeds the one- and two-quadrature Shannon capacities for any average signal power. Although this scenario is highly hypothetical due to rather unrealistic technical requirements, it indicates that there are instances when the Shannon formula does not specify the ultimate capacity of optical communication under the average power constraint. In order to identify the ultimate quantum capacity limits, one needs to describe the input states of light, their propagation, as well as the detection of the optical signal using the mathematical formalism of quantum mechanics. Presenting this formalism in  full detail would go beyond the scope of the present tutorial paper. We will give here only a brief, few-paragraph summary, referring an interested reader to one of excellent textbooks \cite{Nielsen2000, Desurvire2009, Wilde2017}.

Fock states used in the communication scenario proposed by Gordon cannot be legitimately described within the classical theory of electromagnetic radiation. In order to take into account Fock states and any other non-classical states of light, the complex amplitudes $\alpha$ representing the electromagnetic field in individual slots need to be replaced by more intricate mathematical objects, namely density operators, denoted often with a carret as $\hat{\varrho}$ and represented by infinitely dimensional, hermitian, positive semidefinite matrices with a unit trace. The counterpart of well-defined complex field amplitudes is the class of coherent states \cite{GlauberPhysRev1963b,SudarshanPRL1963}. In the quantum mechanical picture of communication also shown in Fig.~\ref{Fig:CommDiagram}, the value $x$ of the input random variable $X$ determines the quantum state $\hat{\varrho}_x$
of the electromagnetic field in a given slot. Propagation through the physical medium is described by a certain map $\hat{\varrho}_x \rightarrow \hat{\varrho}_x'$ acting within the set of density operators. In our case, this map is a generalization of the transformation given in Eq.~(\ref{Eq:alphajchannel}). After propagation, the measurement of the received optical signal produces outcomes described by a random variable $Y$.
The conditional probability distributions $p_{y|x}$ of measurement outcomes $y$ when the field arrives at the receiver in a state $\hat{\varrho}_x'$ can be found using Born's rule. This enables one to calculate the mutual information according to Eq.~(\ref{Eq:IXY}).

The ultimate quantum mechanical capacity limit is obtained by optimizing the mutual information ${\sf I}(X;Y)$ in two domains. The first one involves optimization over all measurements---even hypothetical---that can be performed on the received quantum systems. This task is greatly simplified by Holevo's theorem \cite{Holevo1973},
which states that for any physically permissible measurement one has
\begin{equation}
\label{Eq:Holevoquantity}
{\sf I}(X;Y) \le \chi =  {\sf S} \left( \sum_{x} p_x
 \hat{\varrho}_x' \right)
- \sum_{x} p_x{\sf S}
(\hat{\varrho}_x'),
\end{equation}
where  ${\sf S}(\hat{\varrho}) = - {\text{Tr}} (\hat{\varrho} \log_2
\hat{\varrho} )$ is the von Neumann entropy of a density operator $\hat{\varrho}$. The Holevo quantity $\chi$ defined above has formal structure analogous to mutual information in Eq.~(\ref{Eq:IXY}). The first term is the von Neumann entropy of the average output quantum state after propagation, while the second term is the average von Neumann entropy of an individual output state whose preparation is known.

In the second step, the Holevo quantity $\chi$ needs to be optimized over all ensembles of input quantum states $\hat{\varrho}_x$ with respective probabilities $p_x$ that
satisfy relevant constraints, in the case considered here an upper bound on the average optical power. A rigorous mathematical proof of the quantum limit for the AWGN propagation model has been presented only recently \cite{GiovannettiGarciaPatronNPH2014}. The result confirmed the previously conjectured expression
in the form:
\begin{equation}
\label{Eq:CHol}
{\sf C}_{\text{H}} = g(\ns+ \nn) - g(\nn),
\end{equation}
where $g(\upsilon)$ has been defined in Eq.~(\ref{Eq:g(x)}). In the following, ${\sf C}_{\text{H}}$ will be referred to as the Holevo capacity limit. Interestingly, for any transmission coefficient $\PTr$ and the excess noise value $\nn$, the Holevo quantity $\chi$ calculated according to Eq.~(\ref{Eq:Holevoquantity}) for a continuous Gaussian ensemble of coherent states with complex amplitudes $\alpha \sim {\cal CN} (0,\bar{n})$ yields the capacity limit ${\sf C}_{\text{H}}$. This is a direct counterpart of the input probability distribution saturating the two-quadrature Shannon capacity limit. However, the detection strategy that would achieve the Holevo quantity for this input ensemble remains highly elusive. It has been demonstrated that the inequality in (\ref{Eq:Holevoquantity}) is tight \cite{Hausladen1996,Schumacher1997,Holevo1998}, but the argument used in the mathematical proof cannot be translated in a straightforward manner into feasible detection schemes for optical fields. This is somewhat analogous to the canonical proof of the Shannon noisy channel coding theorem based on the statistics of random codes, which does not necessarily provide a constructive recipe to devise practical error correction algorithms.

It is insightful to compare the Holevo capacity limit with the Shannon capacity limit in specialized parameter regimes. In the absence of excess noise, when $\nn =0$, one has ${\sf C}_{\text{H}} = g(\ns)$. Consequently, the curve shown in Fig.~\ref{Fig:Capacities} as ${\sf C}_{\text{Fock}}$ depicts also more generally  the Holevo capacity limit for loss-only propagation \cite{GiovannettiGuhaPRL2004}. Furthermore, for large average received signal photon number per slot, $\ns \gg 1$, one obtains the following power series expansion in $\ns^{-1}$:
\begin{equation}
\label{Eq:gnsexp}
g(\ns) = \log_2 (1+\ns)
+ \log_2 \Eul - \frac{\log_2 \Eul }{2\ns}
+ O(\ns^{-2}).
\end{equation}
The leading-order term is simply the Shannon capacity of two-quadrature communication derived in Eq.~(\ref{Eq:CS2}) in the special case when there is no excess noise, $\nn=0$. The second-to-leading term specifies the capacity advantage compared to the Shannon limit when $\ns \gg 1$. This advantage is equal to $1~\text{nat} = \log_2 \Eul \approx 1.44~\text{bits}$ of information per slot.
On the other hand, when  the excess noise photon number per slot is much greater than one, $\nn \gg 1$, applying expansion (\ref{Eq:gnsexp}) to both $g(\ns+\nn)$ and $g(\nn)$ yields
\begin{multline}
{\sf C}_{\text{H}} = g(\ns+\nn)-g(\nn) \\
= {\sf C}_{\text{S2}} + \left( \frac{1}{\nn} - \frac{1}{\ns + \nn}\right) \log_2 \Eul + O(\nn^{-2}) .
\end{multline}
It is seen that the $\log_2 \Eul$ terms cancel and the difference between the Holevo capacity and the Shannon capacity becomes minute.

The above results indicate that the Holevo advantage is negligible in scenarios where optical amplification, inevitably generating substantial amounts of excess noise, is used to regenerate propagating optical signals
\cite{AntonelliMecozziJLT2014,JarzynaECOC2019}.
Unconventional communication strategies can be beneficial for short-haul loss-only links, such as optical interconnects. A newly emerging application area may be continuous-variable quantum key distribution (QKD)
\cite{SilberhornRalphPRL2002,GrosshansNATURE2003}. In QKD protocols, generation of a secure cryptographic key requires that the mutual information between the the sender and the receiver exceeds the information about the transmitted signal or measurement outcomes that could be gained by an eavesdropper with unlimited technological capabilities \cite{JouguetKunzPRA2012}. Increasing the mutual information between the legitimate users could improve the key rates or even enable key generation over longer distances. Another potential use case emerges in scenarios where signal regeneration is fundamentally not possible, such as optical communication in space \cite{HemmatiBiswasProcIEEE2011}.

\section{Photon-starved communication}
\label{Sec:PhotonStarved}

The difference between the Shannon and the Holevo capacity limits is most strongly pronounced in the photon-starved regime, when the average received number of signal photons per slot is much less than one, $\ns \ll 1$. This scenario, encountered e.g.\ in deep-space optical communication
\cite{HemmatiWiley2005,BiswasICSOS2017,SodnikICSOS2017,BorosonSPIE2018},
can be viewed as an extreme version of power-limited communication, when received signal power $\PTr\SigPower$ is restricted but the utilized bandwidth is so high that
$\PTr\SigPower/B \ll h f_{\text{c}}$.
In this parameter regime it is convenient to express the maximum attainable information rate defined in Eq.~(\ref{Eq:InfRate}) as
\begin{equation}
\label{Eq:RatePIE}
{\sf R} = B\cdot \ns \cdot {\sf PIE} = \frac{\PTr \SigPower}{h f_{\text{c}}} \cdot {\sf PIE},
\end{equation}
where ${\sf PIE} = {\sf C}/\ns$ is the  photon information efficiency (PIE) specifying how much information is retrieved from one received photon \cite{DolinarICSOS2011}. The product $B\cdot \ns = \PTr \SigPower/(h f_{\text{c}})$ is the number of signal photons received in unit time. As a side note, PIE is closely related to  the information theoretic concept of the capacity per unit cost which has been analyzed within the classical \cite{Verdu1990} as well as the quantum mechanical \cite{JarzynaPRA2017,DingPavlichinIEEETIT2019} framework.

When $n_s \ll 1$, expansion of one- and two-quadrature Shannon capacity limits derived respectively in Eqs.~(\ref{Eq:CS1}) and (\ref{Eq:CS2}) up to the linear term in $n_s$ yields the following expressions for PIE:
\begin{equation}
\label{Eq:PIES1PIES2}
{\sf PIE}_{\text{S1}} \approx \frac{2}{1+2 \nn} \log_2\Eul,
\quad {\sf PIE}_{\text{S2}} \approx \frac{1}{1+\nn} \log_2\Eul.
\end{equation}
When the excess noise is low, $\nn \ll 1$, homodyne shot noise dominates the denominator in both expressions and the PIE is effectively equal to $2~\text{nats} \approx 2.88~\text{bits}$ per photon for one-quadrature communication and $1~\text{nat} \approx 1.44~\text{bits}$ per photon for two-quadrature communication.
For high excess noise, $\nn \gg 1$, both expressions in Eq.~(\ref{Eq:PIES1PIES2}) coincide and are equal to
$(\log_2\Eul)/\nn$. Reverting to the information rate according to Eq.~(\ref{Eq:RatePIE}) one obtains the standard formula for power-limited communication in the form ${\mathsf{R}}_{\text S} \approx B(\ns/\nn) \log_2 \Eul = (\PTr \SigPower/\NoisePSD) \log_2 \Eul$.

\begin{figure}
	\includegraphics[width=\linewidth]{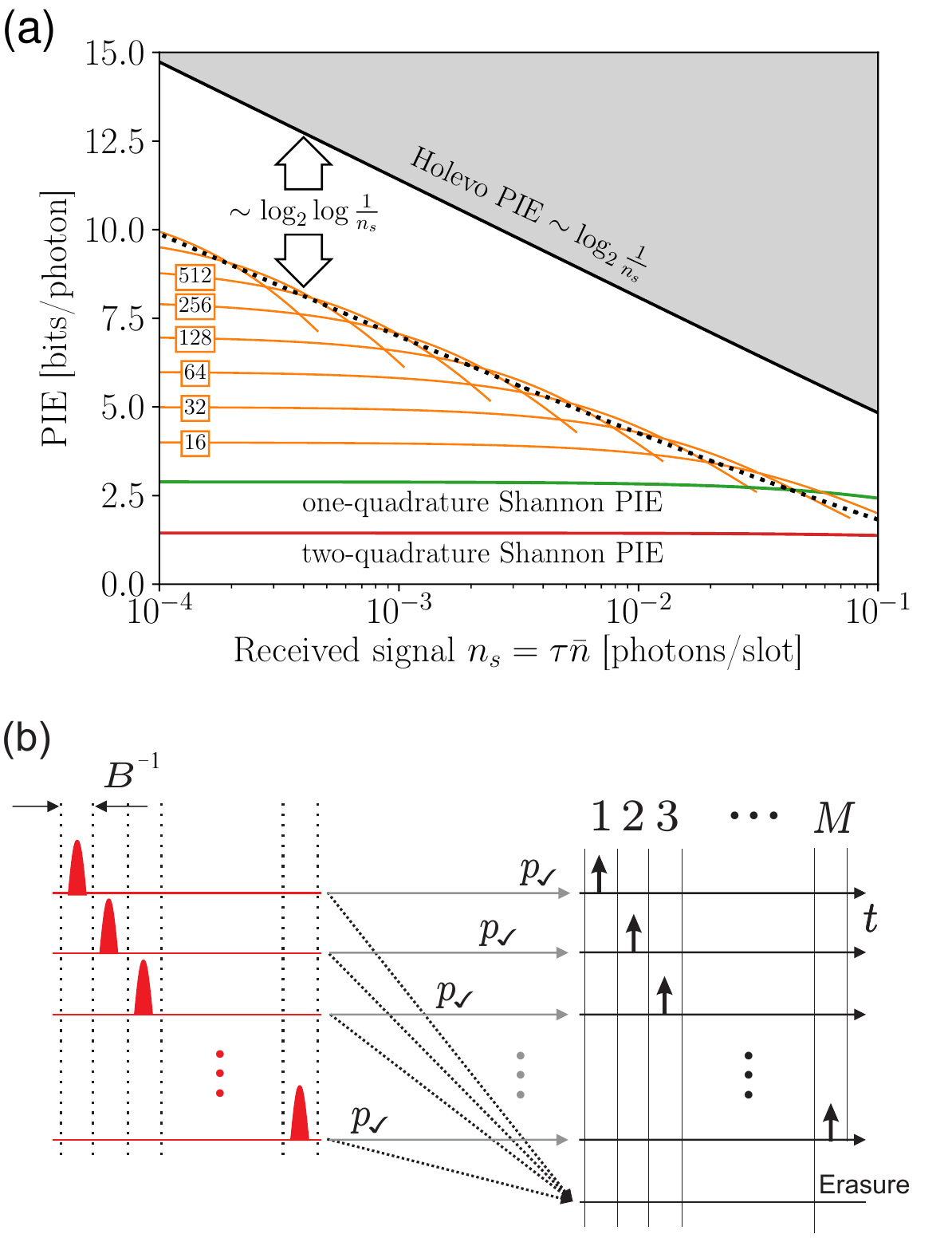}
\caption{(a) The photon information efficiency implied by the one- and two-quadrature Shannon limits compared with the Holevo limit as a function of the average detected photon number $n_s$. Thin lines depict PIE for the directly detected PPM format with the PPM frame length (format order) specified in boxes. The black dotted line represents the approximation derived in Eq.~(\ref{Eq:PIEPPMOpt}). (b) $M$-ary PPM format with direct detection.}
\label{Fig:PIE}
\end{figure}

The above result is in stark contrast with the PIE obtained from the Holevo capacity limit. Consider first loss-only propagation with $\nn=0$. For $\ns \ll 1$ the Holevo capacity limit ${\sf C}_{\text{H}}=g(n_s)$ can be written as a sum of a logarithmic term and a remainder admitting a power series expansion in $n_s$, which yields:
\begin{equation}
\label{Eq:PIEH}
{\sf PIE}_{\text{H}} = \frac{g(\ns)}{\ns} = \log_2 \frac{1}{\ns} + \log_2 \Eul + O(\ns).
\end{equation}
As illustrated in Fig.~\ref{Fig:PIE}(a), the above expression exhibits a qualitatively different scaling with $\ns$ compared to the Shannon limit and it can attain an arbitrarily high value with diminishing $n_s$.

A practical way to achieve high photon information efficiency in optical communications is to use the pulse position modulation (PPM) format combined with direct detection. As depicted in Fig.~\ref{Fig:PIE}(b), $M$-ary PPM format uses $M$ equiprobable multislot symbols defined by the location of one pulse within a frame of $M$ otherwise empty temporal slots. Thus one PPM frame can encode $\log_2 M$ bits of information. In order to ensure a fair comparison with other communication scenarios, the duration of a single slot will be kept at $B^{-1}$. Hence time required to transmit a single PPM symbol is $\PPMtime = MB^{-1}$. Under the average power constraint, the pulse carries the optical energy of the entire frame, equal to $Mn_s$ after transmission. In the absence of excess noise, ideal direct detection allows one to recover the input symbol from the timing of the photodetection event, provided that at least one photocount has been registered.
According to Eq.~(\ref{Eq:pkDD}) specialized to the present scenario,
the probability of such an event is equal to $p_\checkmark  = \sum_{k=1}^\infty p_k = 1 - \exp(-Mn_s)$.
When no photocount is produced over the entire PPM frame, information about the transmitted symbol is erased. The mutual information per slot for such an $M$-ary erasure communication channel is a product of three factors
$ {\mathsf{I}}_{\text{PPM}}^{(M)} =  M^{-1} \cdot p_\checkmark \cdot \log_2 M$
corresponding respectively to renormalization to one temporal slot, the probability that the  erasure has not taken place, and the number of bits encoded in one PPM frame \cite{WasedaSasakiJOCN2011}.
The resulting PIE is depicted in Fig.~\ref{Fig:PIE}(a) for PPM orders that are integer powers of $2$.
For a given PPM order $M$, in the limit $\ns \rightarrow 0$ the PIE approaches the value
\begin{equation}
\label{Eq:PIEPPMM}
{\mathsf{PIE}}_{\text{PPM}}^{(M)} =  \frac{1}{\ns} {\mathsf{I}}_{\text{PPM}}^{(M)} = \frac{1}{M\ns} \cdot p_\checkmark \cdot \log_2 M \rightarrow \log_2 M
\end{equation}
which follows from the linear approximation $p_\checkmark \approx M\ns$. This approximation requires that $M\ns \ll 1$.

When the average received signal photon number per slot $\ns \ll 1$ is fixed, one can identify the optimal PPM order by expanding $p_\checkmark$ up to the quadratic term
\begin{equation}
p_\checkmark \approx M\ns - \frac{1}{2}(M\ns)^2
\end{equation}
and inserting the result into the expression for ${\mathsf{PIE}}_{\text{PPM}}^{(M)}$ given in Eq.~(\ref{Eq:PIEPPMM}). Equating to zero the derivative of the resulting approximate ${\mathsf{PIE}}_{\text{PPM}}^{(M)}$ with respect to $M$,
treated as a continuous real parameter, yields a closed expression for the optimal PPM order $M^\ast$ in the form \cite{JarzynaKuszajOPEX2015}
\begin{equation}
M^\ast \approx \frac{2}{n_s} \left[
W \left( \frac{2\Eul}{n_s}\right)
\right]^{-1},
\end{equation}
where $W(\cdot)$ is the Lambert function \cite{Corless1996} defined by the transcendental equation $W(\upsilon) \Eul^{W(\upsilon)} = \upsilon$. The corresponding optimal PIE value can be written as \cite{JarzynaBanaszekICSOS2017}
\begin{equation}
\label{Eq:PIEPPMOpt}
{\sf PIE}_{\text{PPM}}^\ast \approx \left( W \left( \frac{2\Eul}{n_s} \right) - 2
+ \left[ W \left( \frac{2\Eul}{n_s}\right) \right]^{-1} \right) \log_2 \Eul.
\end{equation}
As seen in Fig.~\ref{Fig:PIE}(a), this expression slightly underestimates the optimal value of PIE.
For large arguments $\upsilon\gg 1$ the Lambert function admits expansion $W(\upsilon) = \log \upsilon - \log\log \upsilon + o(1)$. Using this expansion in Eq.~(\ref{Eq:PIEPPMOpt}) and comparing the result with the Holevo PIE  calculated in Eq.~(\ref{Eq:PIEH}) reveals a gap between the PIE of the optimized PPM format with direct detection on one hand and the ultimate quantum limit on the other hand \cite{KochmanWangTIT2014}. This gap is characterized in the leading order by a double-logarithmic term of the form $\log_2 \log (1/n_s)$.

\begin{figure}
	\includegraphics[width=\linewidth]{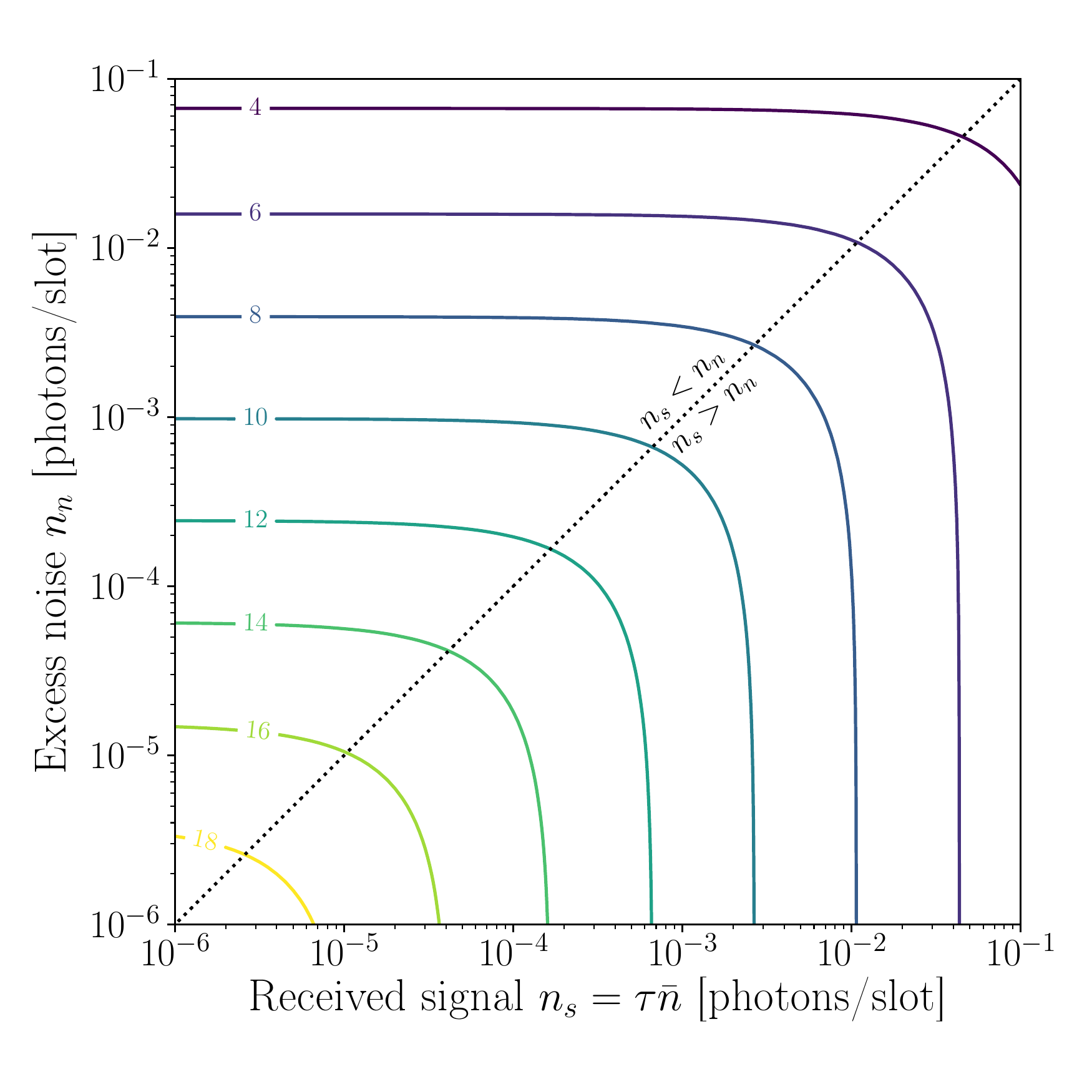}
\caption{The photon information efficiency calculated from the Holevo capacity limit as a function of the average signal $\ns$ and excess noise $\nn$ photon numbers per slot.}
\label{Fig:PIEHnsnn}
\end{figure}

In practice, the propagating optical signal will always acquire some excess noise, contributed e.g.\ by scattered stray light. Its impact can be estimated using
Fig.~\ref{Fig:PIEHnsnn}, which depicts the Holevo limit on the photon information efficiency $\mathsf{PIE}_{\text{H}} = \mathsf{C}_{\text{H}}/\ns$ as a function of the signal $\ns$ and the excess noise $\nn$ photon numbers per slot. It is seen that the noiseless analysis holds as long as $\nn \ll \ns$. For a fixed non-zero excess noise figure $\nn$, the PIE remains finite with the maximum value attained when $\ns \ll \nn$:
\begin{equation}
\mathsf{PIE}_{\text{H}} = \frac{1}{n_s} \bigl( g(\ns + \nn) - g(\nn)\bigr)
\stackrel[\ns \rightarrow 0]{}{\longrightarrow}  \log_2 (1+ \nn^{-1}).
\end{equation}
Thus the general Holevo limit on the maximum attainable information rate in the power-limited regime with unrestricted bandwidth takes the form
\begin{equation}
{\mathsf{R}}_{\text{H}} \approx B \cdot \ns \cdot \log_2 (1+ \nn^{-1}) = \frac{\PTr\SigPower}{h f_c} \log_2 \left( 1+ \frac{h f_c}{\NoisePSD} \right).
\end{equation}
Notably, the second expression, involving dimensional physical quantities, depends explicitly on the energy $h f_{\text{c}}$ of a single photon at the carrier frequency. This energy defines the absolute scale for the noise power spectral density below which the quantum nature of light starts to play a non-trivial role. Only when
$\NoisePSD \gg h f_c $ one can expand the logarithm into a power series to obtain the Shannon expression ${\mathsf{R}}_{\text{S}} \approx (\PTr \SigPower / \NoisePSD) \log_2 \Eul$.

In the model considered above only excess noise added to the signal wavepacket profile  has been taken into account in accordance with Eq.~(\ref{Eq:alphajchannel}). When standard direct detection is used, one should include in the analysis excess noise present in the entire time-bandwidth area measured by the photodetector \cite{BorosonNoiseSPIE2018}. In the basic model for such a scenario,
when the time-bandwidth area detected per slot is much larger than one,
the effective statistics of background counts generated by the excess noise can be described by Poissonian distribution \cite{GagliardiKarp1995}. The photon information efficiency of such a noisy PPM link can be
analyzed using a relative entropy bound \cite{HamkinsISIT2004}. If the photodetector discriminates only between zero and at least one photocount in each slot, the dependence of PIE on the signal and the noise strengths has a qualitatively similar character to that shown in Fig.~\ref{Fig:PIEHnsnn}
\cite{ZwolinskiJarzynaOpEx2018,BanaszekZwolinskiICSOS2019}. It is worth noting that the technique of quantum pulse gating \cite{BrechtReddyPRX2015,AllgaierAnsariNCOMM2017,ReddyRaymerOPTICA2018} can be used as a noise-rejection mechanism for the received optical signal that potentially has both unit efficiency and unit selectivity \cite{ShahverdiSuaSciRep2017}. This technique combined with photon number resolving photodetection in principle could allow one to approach the Holevo limit in photon-starved communication \cite{BanaszekSPIE2019}.

\section{Joint multisymbol detection}
\label{Sec:Multisymbol}

\begin{figure}
	\includegraphics[width=\linewidth]{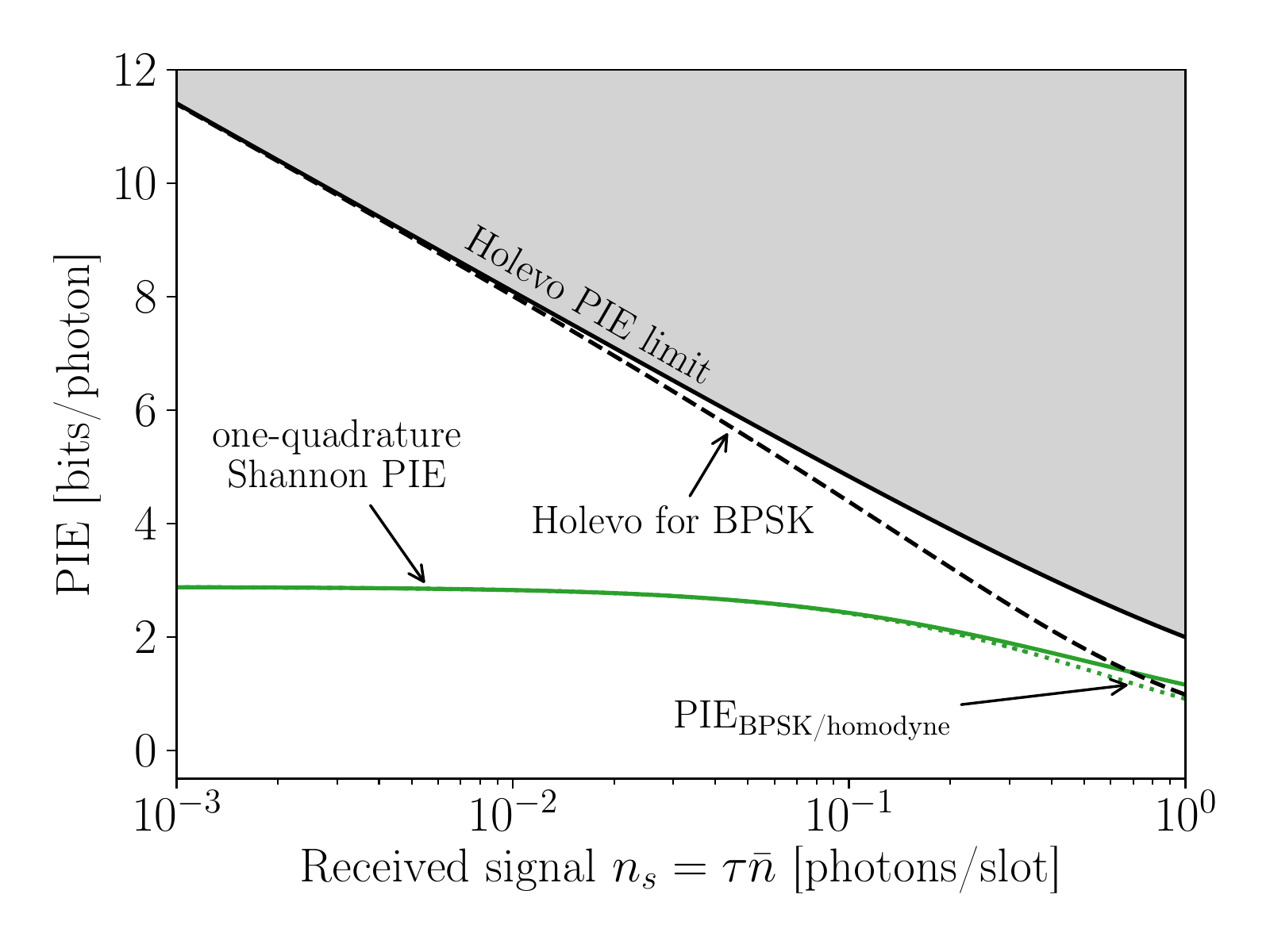}
\caption{The photon information efficiency (PIE) calculated from the one-quadrature Shannon capacity limit and from the Holevo capacity limit (solid lines) compared with the photon information efficiency for the BPSK constellation assuming homodyne detection and general physically permissible detection strategies included in the Holevo quantity (dashed lines).}
\label{Fig:BPSK}
\end{figure}

As pointed out in Sec.~\ref{Sec:Holevo}, the Holevo theorem does not provide a systematic way to design practical measurements that saturate the Holevo quantity for a given input ensemble of quantum states. Nevertheless, it can motivate search for detection strategies that go beyond conventional approaches.
As a simple example, consider the BPSK constellation, represented in the quantum mechanical formalism by two equiprobable coherent states with the same mean photon number and phases $0^\circ$ and $180^\circ$. Loss-only propagation attenuates their amplitudes to $\pm\alpha'$, where $\alpha' = \sqrt{\ns}$. In the photon-starved regime, when $\ns \ll 1$, shot-noise-level homodyne detection of the $I$ quadrature yields PIE that practically overlaps with that implied by the one-quadrature Shannon capacity limit, as shown in Fig.~\ref{Fig:BPSK}. In contrast, the Holevo quantity $\chi_{\text{BPSK}}$ calculated for the BPSK constellation
yields photon information efficiency that is very close to the Holevo capacity limit. This result indicates that photon-efficient communication can be in principle achieved with the BPSK constellation, but conventional homodyning needs to be replaced by another detection strategy.

In general, two prerequisites are required to saturate the Holevo quantity. The first one is that quantum states drawn from the input ensemble are assembled into words transmitted over multiple channel uses (i.e.\ many temporal slots in the optical scenario discussed here). This is a straightforward analog of classical encoding. However, the second assumption is that collective measurements are performed on blocks of received elementary quantum systems that carry the entire words. Such joint detection strategies can be much more powerful than measurements performed individually on received quantum systems. This is intimately related to the fact that any quantum measurement reveals only partial information about the measured physical system.

\begin{figure}[t]
	\begin{center}
		\includegraphics[scale=0.95]{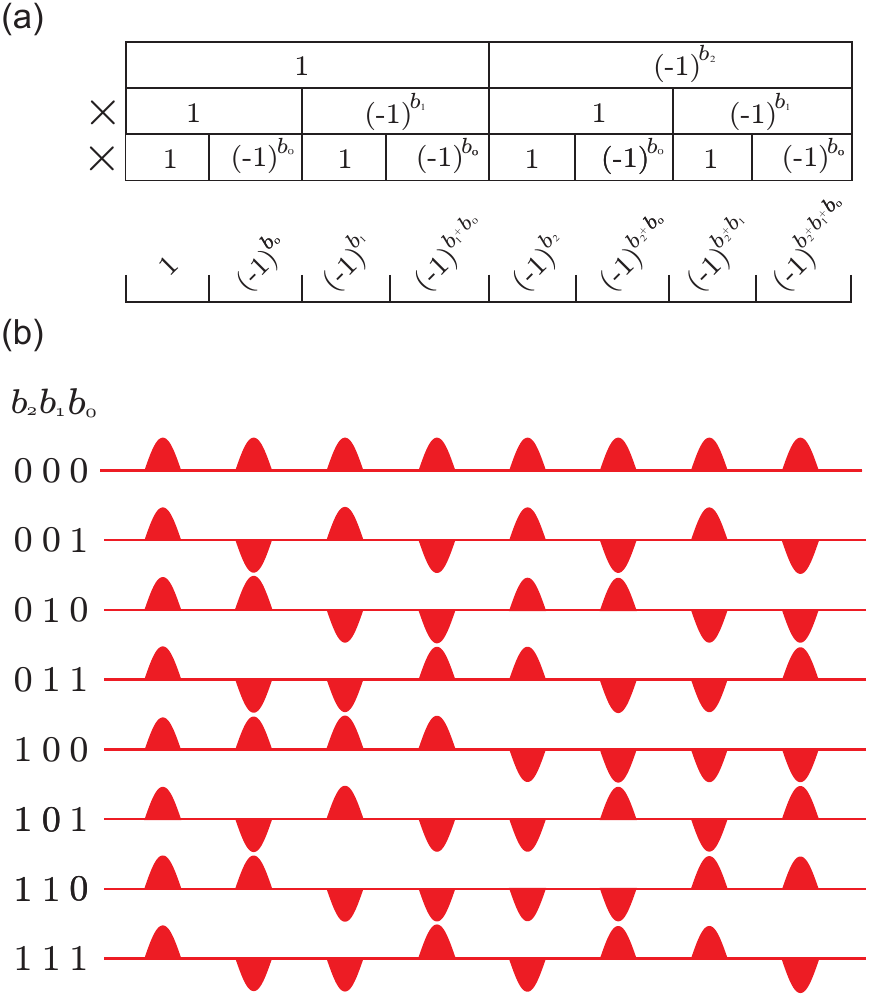}
		\caption{(a) The construction of Hadamard  words of length $M = 2^m =8$. For the $l$th word, $l=1,2,\ldots, M$, the integer $l-1$ is expressed in the binary representation by a bit string $b_{m-1} b_{m-2}\ldots b_1 b_0$ which defines a hierarchy of phase factors shown in the diagram. Vertical multiplication of the phase factors along columns yields a Hadamard word corresponding to a given $l$. (b) The recipe applied to the construction of Hadamard words of length $M = 8$, depicted as sequences of optical pulses pointing up  for the `$+$' phase factor and  pointing down for the `$-$'factor, labelled with the corresponding bit strings
 $b_{2} b_{1} b_{0}$.} \label{Fig:HadamardPhase}
	\end{center}
\end{figure}

The above aspects can be illustrated with a very elegant communication strategy utilizing the BPSK format that
has been described by Guha \cite{GuhaPRL2011}. The basic idea is to transmit words composed from BPSK symbols defined by rows of a Hadamard matrix. We will refer to these sequences as {\em Hadamard words}. Hadamard matrices are real orthogonal matrices with entries $\pm 1$ and exist for dimensions $M=2^m$ that are integer powers of 2. The construction of Hadamard words for $M=8$ is shown graphically in Fig.~\ref{Fig:HadamardPhase}(a). The starting point to find the $l$th  Hadamard word of length $M$, $l=1,2,\ldots, M$, is to write $l-1$ in the binary representation using an $m$-bit string $b_{m-1} b_{m-2}\ldots b_1 b_0$ so that $l-1=\sum_{i=0}^{m-1} 2^i b_i$. The $i$th bit contributes a multiplicative phase factor alternating between $1$ and $(-1)^{b_i}$ every $2^i$ positions. Individual entries in the $l$th Hadamard word are products of all these $m$ factors and determine phases of BPSK symbols in the corresponding Hadamard word, as depicted in Fig.~\ref{Fig:HadamardPhase}(b).

\begin{figure}[t]
	\begin{center}
		\includegraphics[scale=1.1]{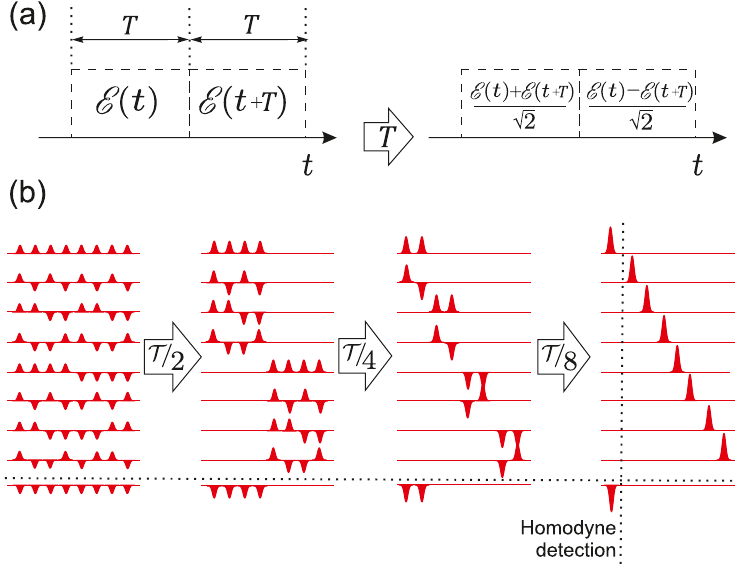}
		\caption{(a) A time-domain interferometer superposes the optical field in two adjacent time intervals  $T$. (b) The cascade of time domain interferometers which implements all-optical mapping of BPSK Hadamard words of duration $\PPMtime = MB^{-1}$ onto the PPM format shown for $M=8$. The dashed lines graphically separate a modification that consists in adding the $--\ldots-$ word and performing homodyne detection in the first temporal slot of the output from the cascade.} \label{Fig:HadamardPulses}
	\end{center}
\end{figure}

The essence of the joint detection strategy for BPSK Hadamard words is to use optical interference to concentrate the optical energy of the entire word in a location that is different for each input word. This goal can be achieved using a cascade of interferometric modules \cite{BanaszekICSOS2017}. As shown schematically in Fig.~\ref{Fig:HadamardPulses}(a), one module superposes the optical field in two adjacent time intervals of duration $T$. Because of the mathematical construction of Hadamard words described above, sending the pulse sequence defined by the $l$th word through a cascade of interferometers that operate on time intervals that correspond to one half, one quarter, etc.\ fractions of the word duration $\PPMtime = MB^{-1}$ down to a single temporal slot $B^{-1}$, concentrates the entire optical energy in the $l$th temporal slot at the output of the cascade, as depicted in Fig.~\ref{Fig:HadamardPulses}(b). If ideal, shot-noise-level direct detection is implemented at this output, the information efficiency is equivalent to that of an $M$-ary PPM link analyzed in Sec.~\ref{Sec:PhotonStarved}.
An interesting feature of communication using BPSK Hadamard words is that high PIE is achieved with optical power uniformly distributed across temporal slots, which is in stark contrast with the PPM format. In the latter case increasing PIE requires generating single pulses within frames covering a larger number of temporal slots. This results in a demand for the increasing peak-to-average power ratio of the optical PPM signal, which may be constrained by the physics of the transmitter laser system. In the case of BPSK Hadamard words, the effective format order is increased by changing the phase modulation pattern. The drawback is  a much more complex interferometric receiver whose construction depends on the format order.

The fact that detection of individual symbols and postprocessing of measurement outcomes is usually insufficient to saturate the Holevo capacity limit is related to a phenomenon known in quantum information theory as the superadditivity of accessible information \cite{SasakiKatoPRA1998,BuckvanEnkPRA2000,ChungIEEETIT2016}. In the case of the BPSK constellation, it can be shown that no physically permissible measurement on individual symbols can beat the PIE limit of $2\log_2 \Eul \approx 2.88$~\text{bits/photon} that is achieved with conventional homodyne detection in the photon-starved regime. Communication with jointly detected BPSK Hadamard words described above can be viewed as an illustration of the superaddivity phenomenon when a collective measurement is performed on at least $M=8$ symbols, as then PIE achieves $\log_2 8 = 3$~bits/photon for $\ns \ll 1$. Superadditivity of accessible information can be also demonstrated with measurements on fewer than eight phase shift keyed symbols
\cite{GuhaPRL2011,KlimekJachuraJMO2016,RosatiMariPRA2016,KunzXXX2019}. As an example also shown in
Fig.~\ref{Fig:HadamardPulses}(b), consider the set of $M$-ary BPSK Hadamard words enlarged by adding an $(M+1)$st sequence $--\ldots-$. The sequences
$++\ldots+$ and $--\ldots-$
are sent with  probabilities $p_1/2$ each, while the remaining $M-1$ Hadamard words are used with the same probability $(1-p_1)/(M-1)$. At the output of the interferometric cascade shown in Fig.~\ref{Fig:HadamardPulses}(b) direct detection is performed in all temporal slots except the first one where the optical energy of the sequences $++\ldots+$ and $--\ldots-$ becomes concentrated. In this slot, homodyning is used to measure the $I$ quadrature. The complete detection outcome consists of the continuous quadrature value for the first slot and a discrete variable specifying in which slot, if any, a photocount has occurred. Optimizing mutual information with respect to $p_1$ yields for $M=2,4$, and $8$ in the limit $n_s \ll 1$ the respective values of the photon information efficiency $\mathsf{PIE} = 2.98, 3.10$, and $3.39$  bits/photon. These figures exceed the Shannon limit as well as the performance of the directly detected PPM format.

\section{Conclusions}
\label{Sec:Conclusions}

The purpose of this tutorial paper was to provide an elementary introduction to quantum mechanical capacity limits of optical communication links. The discussion was based on an elementary model of a narrowband optical signal acquiring excess additive white Gaussian noise in the course of propagation. The crucial issue is the comparison between the excess noise power spectral density $\NoisePSD$ and the energy $hf_{\text{c}}$ of a single photon at the carrier frequency $f_{\text{c}}$ per unit time-bandwidth area. When $\NoisePSD \gg hf_{\text{c}}$, the standard Shannon capacity limit for conventional quadrature measurements is applicable and the performance of a communication link can be characterized in terms of the signal-to-noise ratio.

The situation becomes more nuanced when $\NoisePSD \ll hf_{\text{c}}$. In this regime the particle nature of light plays a non-trivial role and the energy of a single photon at the carrier frequency defines the absolute scale for quantifying the signal and the noise strengths. In the discrete slot model used in this tutorial, two relevant figures of merit are the average number of signal $\ns$ and noise $\nn$ photons per slot. The ultimate capacity limit given in Eq.~(\ref{Eq:CHol}) follows from Holevo's theorem and it depends explicitly on both $\ns$ and $\nn$ rather than their ratio. This reflects the fact that the Holevo capacity limit involves optimization over all physically permissible detection strategies for which no single universal noise figure can be defined. When the average signal photon number per slot significantly exceeds one, $\ns \gg 1$, the advantage of the Holevo capacity limit is $1~\text{nat} = \log_2 \Eul \approx 1.44$~bits per slot compared to the Shannon limit. So far not much is known about practical designs for receivers that would beat the Shannon limit in this case. In the photon-starved regime, when $\ns \ll 1$, photon counting detection of intensity-modulated signals can approach the Holevo capacity limit in the leading order, as exemplified by the PPM format optimized with respect to the frame length.

Many interesting questions arise regarding quantum capacity limits beyond the elementary linear AWGN model considered here. Examples include quantum effects in non-linear signal propagation \cite{CorneyHeersinkPRA2008,KunzParisJOSAB2018}
and unconventional communication strategies in the presence of non-Gaussian noise
\cite{TrapaniTekluPRA2015,DiMarioKunznpjQI2019}.
Also, adopting a more general perspective on the time-frequency structure of optical signals may inspire novel modulation formats and receiver designs \cite{BurenkovTikhonovaOPTICA2018,BanaszekJachuraICSO2018}.

\section*{Acknowledgment}

Insightful discussions with Christian Antonelli, Ren\'{e}-Jean Essiambre, Saikat Guha, Gerhard Kramer,
Christoph Marquardt, Antonio Mecozzi, Mark Shtaif, as well as our collaborators on research projects related to quantum aspects of optical communications, are gratefully acknowledged.

\bibliographystyle{myIEEEtran}
\bibliography{limits}

\end{document}